# High temperature mechanical properties and microstructure of hard TaSiN coatings

M.A. Monclús[1*]; L. Yang[2]; I. López-Cabañas[1,3]; M. Castillo-Rodríguez[1]; A. Zaman[3]; J. Wang[2]; E.I. Meletis[3]; R. Gonzalez-Arrabal[4]; J. Llorca[1,5] and J.M. Molina-Aldareguía[1*]

[1] IMDEA Materials Institute, c/Eric Kandel 2, 28906 Getafe, Madrid, Spain;
[2] China Aerodynamics Research and Development Center, Mianyang, China, 621000;
[3] Department of Materials Science and Engineering, University of Texas at Arlington, TX, 76019, USA;
[4] Instituto de Fusión Nuclear Guillermo Velarde (UPM) and Departamento de Ingeniería Energética (ETSII/UPM), José Gutiérrez Abascal 2, 28006, Madrid, Spain;
[5] Department of Materials Science, Polytechnic University of Madrid, E.T.S. de Ingenieros de Caminos, 28040 Madrid, Spain

* Corresponding authors.

## Abstract

Room and high temperature mechanical properties of reactive magnetron sputtered TaSiN coatings were measured using nanoindentation (between 25°C and 500°C). Fracture toughness was also evaluated at a similar temperature range using the micropillar splitting method. The influence of the nitrogen concentration on the evolving phases and microstructure of the TaSiN coatings, before and after the high temperature testing, were examined by X-ray diffraction (XRD) and transmission electron microscopy (TEM) analysis. XRD spectra showed broad peaks with hexagonal γ-$Ta_2N$ as the main phase, with the cubic δ-TaN phase emerging for higher N contents. Phase composition remained unchanged before and after the 500°C tests. However, after the high temperature tests, TEM analysis showed the presence of an oxide surface layer, with a thickness that decreased (from 42 to 15 nm) with N content, due to residual oxygen diffusion, which replaces nitrogen to form amorphous $SiO_x$. Beneath the oxide-rich surface layer, coatings exhibited a stable nanocrystalline columnar microstructure. Hardness and fracture toughness increased with N content, initially due to the formation of an amorphous Si-N tissue at grain boundaries, and for even higher N contents, due to the appearance of the hard cubic δ-TaN phase. Hardness at 500°C decreased only by 15%, while fracture toughness followed the opposite trend, due to increased plasticity with temperature. The optimum composition turned out to be $Ta_{55}Si_{10}N_{35}$, which retained a hardness of 30 GPa at 500°C, being also the toughest. These observations make this system very interesting for high temperature applications.

## 1. Introduction

Hard nitride coatings are widely used in the cutting tool industry for their favourable mechanical properties and thermal stability [1]. Among all nitride coatings, TaN has gained increasing interest in recent years due to its exceptional properties, such as high hardness [2], good abrasive and wear-resistant properties [3], corrosion resistance [4] and chemical stability [5]. However, polycrystalline TaN has been found to be unstable at temperatures ≥ 500°C because grain boundaries act as fast diffusion paths [6] and also because TaN, particularly if N deficient, oxidizes to $Ta_2O_5$ when annealed in air at temperatures > 600°C [7].

The addition of Si into the TaN coating system can improve the high temperature mechanical properties, since it has been shown to enhance its oxidation resistance [7] by stabilizing their amorphous structure while inhibiting oxygen diffusion. Therefore, the TaSiN system is of great interest for many engineering applications that require protective coatings because of their potential unique combination of high hardness, thermal stability and oxidation resistance [8–10].

The Ta-N system can exhibit a wide variety of metastable and stable phases depending on the deposition conditions [11] and therefore, the microstructure of TaSiN coatings can be quite complex. Some investigations on the thermal stability and mechanical properties of TaSiN coatings with high Si content (>20 at.%) showed that these coatings tend to be mainly amorphous [8,12], while others show that they are mainly nanocrystalline if the Si content is low (<20 at.%) [13]. Zaman *et al.* [13] studied the structural evolution of TaSiN coatings deposited by reactive magnetron sputtering as a function of nitrogen content in the reactive gas. The structure evolved with $N_2$ content in the plasma, from textured hexagonal γ-$Ta_2N$ (7% $N_2$) to a mixture of γ-$Ta_2N$ and fcc δ-TaN (10%-13% $N_2$), and finally to δ-TaN (15% $N_2$). This evolution was accompanied with an increase in the hardness of the coatings from 30 to 40 GPa. Moreover, TaSiN coatings with optimum N content exhibited low wear rates and were thermally stable up to 800°C. However, to the best of our knowledge, there are no reports on the high temperature mechanical properties of TaSiN coatings. The aim of this study was to analyse the effect of N content on the high temperature mechanical properties of TaSiN coatings deposited by reactive sputtering, including fracture toughness, which is a very important parameter for long service lifetime of hard coatings used for a number of applications, such as cutting tools [14], and to study the microstructure evolution of the coatings prior to and after the high temperature testing.

2. Experimental details

TaSiN coatings were deposited in a homemade physical vapour deposition (PVD) system by magnetron co-sputtering from high purity Ta (99.95%) and Si (99.99%) targets in the presence of an Ar+$N_2$ atmosphere on Si <100> single-crystal wafers. The substrate DC bias was -100V to reduce oxygen contamination [15]. The power of the Ta and Si targets was DC-150W and RF-13W, respectively. The total gas pressure ($P_{Ar}+P_{N2}$) and the target-substrate distance were kept constant at 0.66 Pa and 10 cm, respectively, whereas the nitrogen partial pressure was varied from 7% to 15%. The substrate temperature was set to 550ºC. The base pressure of the chamber prior deposition was around 4×$10^{-4}$ Pa. Before deposition, the Ta target and the substrate were sputter cleaned with an Ar plasma. After the cleaning process, a thin layer (~ 40 nm thick) of pure Ta was deposited on the substrate followed by the TaSiN coating deposition, which lasted for one hour. In the rest of the manuscript, the coatings are designated TaSiN-X, where X refers to the nitrogen partial pressure during deposition.

Rutherford backscattering spectroscopy (RBS) was used to determine the elemental composition of the TaSiN coatings, using a He$^+$ beam at the energy of 3.7 MeV in a high vacuum chamber. The backscattered ions were detected by a standard Si-barrier detector located at angles of 165º to the beam direction. At this energy, the nuclear reaction channel $^{14}N(\alpha,p_0)^{17}O$ is open. By coincidence, the energy of the emitted protons is nearly equal to that of the scattered particles and at the resonance energy, the proton and a particle yield could not be separated in our measurements. For this reason, the non-RBS spectra were measured under the same conditions by placing a 13 mm Mylar foil in front of the detector which prevents the α particles from arriving at the detector and allows us to estimate the contribution of the scattered protons to the nitrogen peak [16]. More detailed information about the non-RBS measurement and analysis procedure can be found in [17]. The atomic percentage of Ta, Si and N in the coatings was estimated by comparing experimental and simulated spectra. For the simulations, the commercial computer code SIMNRA was used [18].

The phases present in the coatings were identified using an X-ray diffractometer (XRD, X'Pert PRO MPD, PANalytical) in the θ-2θ configuration using Cu Kα radiation from a Cu anode operated at 45 kV and 40 mA.

The structure of the coatings across the film thickness was examined using a transmission electron microscope (TEM, FEI Talos F200X) operated at 200 kV. Cross-sectional specimens for TEM analysis were produced by ion milling using a FEI Helios Nanolab 600i dual beam FIB-FEGSEM. A Pt layer was deposited to protect the free surface of the sample.

Room and high temperature nanoindentation measurements were performed using a Bruker´s Triboindenter TI950 equipped with a high temperature stage (xSol®) and a Berkovich diamond tip fitted to a long insulating shaft. The sample was placed between two resistive heating elements that eliminate temperature gradients across the sample thickness. Dry air and argon around the tip and sample surface were used to purge the testing area in order to prevent heated gases reaching the transducer and to reduce possible oxidation. Once the sample reached the selected temperature and was stable at that temperature to within ±0.01ºC, the tip was placed at ≈100 μm from the sample surface for 10-15 minutes to ensure passive heating of the tip before the start of the test and to minimize thermal drift. The coatings were tested at 25, 200, 350, 500°C and after cooling to room temperature. At least six indentations were performed at each temperature using load control with loading, holding and unloading times of 10, 5 and 2 seconds respectively and a peak load of 7 mN, which resulted in indentation depths in the range 100-130 nm. No substrate effects were expected as penetration depths were kept below 10% of the total coating thickness [19]. Hardness and reduced elastic modulus were determined using the Oliver-Pharr method [20].

The micropillar splitting method, first proposed by Sebastiani et al. [21], was used to evaluate the micro-scale fracture toughness of the coatings. The method is based on indenting with a sharp tip on top of a micropillar until fracture occurs. The fracture toughness ($K_{IC}$) is determined from the critical load at failure ($P_c$) and the pillar radius ($R$) through:

$$K_{IC} = \gamma \frac{P_c}{R^{3/2}} \qquad (1)$$

where the γ factor depends on the material properties (E/H ratio). The variation of the γ factor as a function of the material parameter E/H for the cube corner geometry has been established by Ghidelli et al. [22] and was used here together with the E and H values of our TaSiN coatings to obtain the γ factor for the different investigated temperatures. Micropillars were machined by focused ion beam (FIB) milling using the FEI Helios Nanolab 600i with a length to diameter aspect ratio of ≈ 1 and a height to film thickness ratio of ≈ 1.0, which eliminates residual stress effects [23]. The micropillars were indented using the Triboindenter TI950 equipped with a diamond cube-corner indenter tip. The system is capable of imaging and performing the tests *in-situ* using a scanning probe microscope (SPM) module. This is critical for the micropillar splitting tests, because the cracks should be initiated and propagated, maintaining symmetry, from the corners of the indent imprint on the top surface of the pillar. High temperature micropillar splitting tests were carried out at 20, 350, 500°C and after cooling to room temperature, using the same hot stage configuration as for the high temperature nanoindentation tests and following the same thermal stabilization approach.

## 3. Results

*3.1. Chemical composition and microstructure of the as-deposited coatings*

According to the RBS measurements, the chemical composition of the TiSiN-7 and TiSiN-10 coatings was homogeneous along their entire thickness, whereas, in the case of the TiSiN-13 and TiSiN-15 coatings, the N content gradually increased with thickness to reach a plateau in N content that remained constant for the top ~500 nm of the coating thickness. Table I gathers the thickness and elemental composition of the deposited TaSiN coatings, considering only a depth of ~500 nm below the surface, where the elemental composition of the coatings was homogeneous in all cases. The nitrogen atomic content increased with the nitrogen partial pressure in the sputtering chamber to a maximum of around 35 at.% for a nitrogen partial pressure of 13%, and remained constant with further increases in $P_{N2}$. A similar trend has been observed for other nitrides [24]. The Ta and Si atomic content decreased accordingly with increasing the N atomic content, but the Si/Ta ratio remained close to 0.2 in all cases.

Table I: Thickness and elemental composition of the deposited TaSiN coatings using nitrogen partial pressures of 7, 10, 13 and 15%.

| Sample | TaSiN-7 | TaSiN-10 | TaSiN-13 | TaSiN-15 |
|---|---|---|---|---|
| N$_2$ fraction in Ar+N$_2$ mixture | 7% | 10% | 13% | 15% |
| Thickness (µm) | 1.00 | 1.25 | 1.26 | 1.46 |
| Film Composition at.% (RBS) | | | | |
| [Ta] | 70 | 60 | 55 | 55 |
| [Si] | 14 | 13 | 10 | 10 |
| [N] | 16 | 27 | 35 | 35 |

As shown in the ternary phase diagram of Ta-Si-N corresponding to the isotherm calculated at 625°C and 1 atm [25], the TaSiN system can form many equilibrium compound phases for the range of compositions of the coatings, namely TaN, Ta$_2$N, Ta$_2$Si, Ta$_5$Si$_3$, TaSi$_2$ and Si$_3$N$_4$, while other metastable phases with different degrees of crystallinity might also be present, especially considering that the deposition conditions during sputtering are far from equilibrium. Therefore, X-ray diffraction (XRD) and transmission electron microscopy (TEM) studies are required to elucidate the phase composition and microstructure of the coatings.

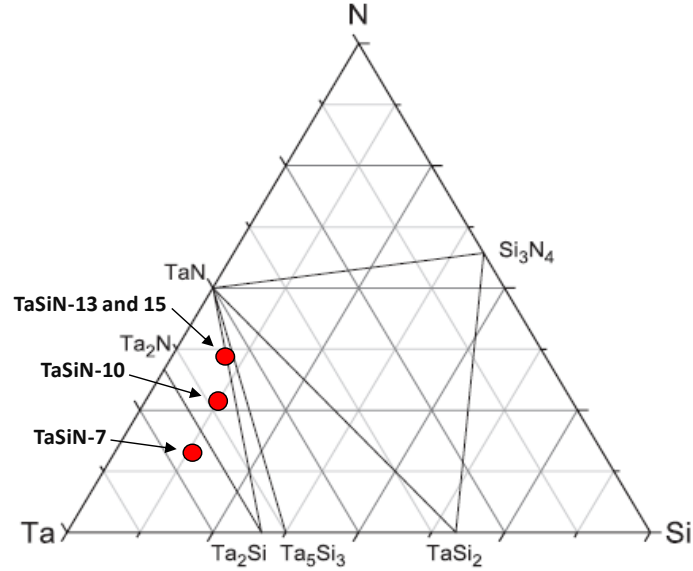

Fig. 1. Phase diagram of ternary Ta–Si–N system, showing the positions of the as deposited TaSiN studied in this work which are represented as circles.

Previous XPS results confirmed the presence of both Ta-N and Si-N containing phases [13]. The XRD patterns of the as-deposited TaSiN coatings are shown in Fig. 2a. The broad peaks in the XRD patterns suggest that the coatings are highly disordered, with very small grains. The XRD pattern of the coating with the lowest N content (TaSiN-7) showed the dominance of a broad peak centred at ≈ 38°, which was attributed to the hexagonal γ-Ta$_2$N phase (PDF# 01-089-4764), with a preferred (101) orientation. However, this peak was slightly shifted to smaller diffraction angles than the theoretical position, implying an increased lattice parameter. With increasing N content, the peak position of this broad peak shifted to match the theoretical position of the γ-Ta$_2$N (101) peak and a new peak emerged at around 34°, corresponding to the (100) plane of the hexagonal γ-Ta$_2$N phase. A third peak at ≈ 35.9° emerged in between the (100) and (101) γ-Ta$_2$N peaks for the coatings with the highest nitrogen content (TaSiN-13 and TaSiN-15), which can be attributed to the (111) plane of the cubic δ-TaN phase (PDF# 04-004-0787). Therefore, the TaSiN-13 and TaSiN-15 coatings consisted of a mixture of hexagonal γ-Ta$_2$N and cubic δ-TaN, probably co-existing with other amorphous phases. The fcc δ-TaN phase is a metastable phase, which frequently appears during sputter deposition [11]. The results are completely consistent with TEM studies carried out before [13], showing that the microstructure of the films deposited at 7% N$_2$ consisted of columnar and highly textured nanocrystalline grains of γ-Ta$_2$N. Increasing the N$_2$ content to 10% reduced the texture and induced the formation of an amorphous Si-N tissue in the grain boundaries, while further increases to 13%, resulted in the appearance of the δ-TaN phase. Fig 2b shows a cross-sectional bright field (BF) TEM image of the TaSiN-15 coating showing evidence of the columnar microstructure, with columns being about 25-40 nm wide. Fig. 2c shows an example of a cross-sectional high resolution (HR) TEM image of the TaSiN-13 coating, showing the nanocrystalline grains and the amorphous Si-N tissue.

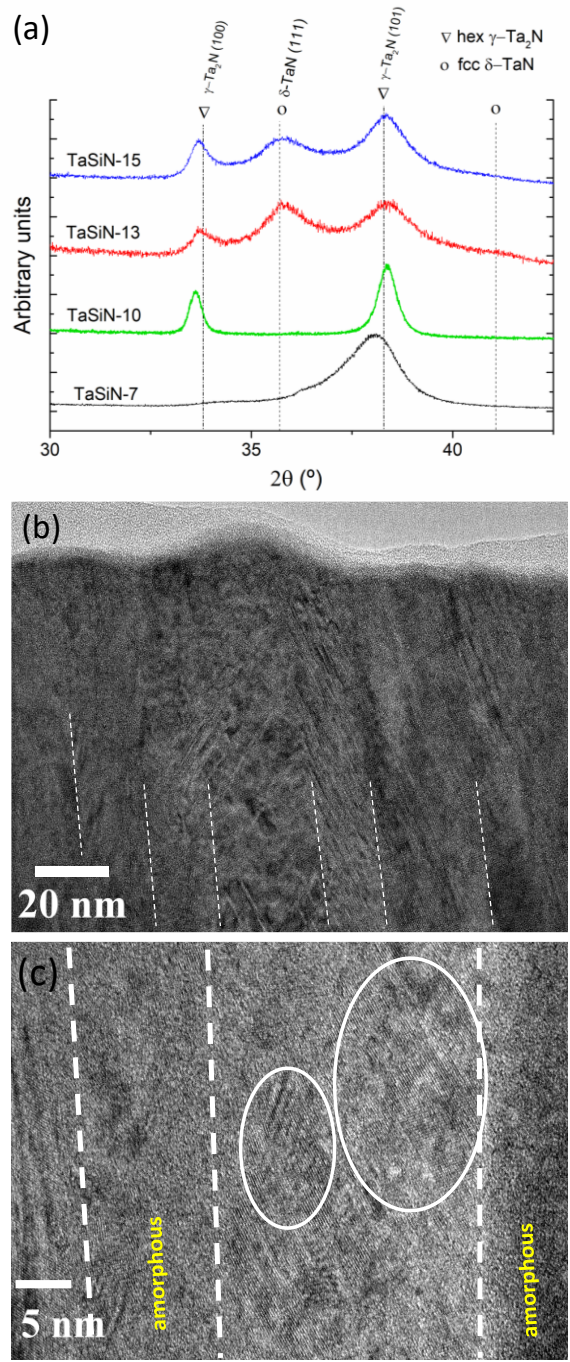

Fig. 2. (a) XRD spectra of the TaSiN films coatings; (b) BF TEM image of TaSiN-15 clearly showing signs of columnar morphology with columns being about 25-40 nm wide; and (c) representative HR TEM image of the microstructure of the TaSiN-13 coating, where columns are composed of several nanograins (circled in white) separated by amorphous regions.

The XRD findings are supported by the RBS results (Table I), from which it was found that the N/Ta atomic ratio increased with $N_2$ partial pressure from 0.23 for TaSiN-7 to 0.63 for TaSiN-13 and TaSiN-15. For TaSiN-7, the N content was not sufficient to form the stoichiometric γ-$Ta_2$N phase, and the fact the (101) γ-Ta$_2$N peak in XRD appeared offset to lower angles, together with the absence of a distinct amorphous phase in the TEM, suggested that the Si might be incorporated into the N deficient γ-Ta$_2$N lattice. For TaSiN-10, however, the N/Ta ratio of 0.45 ensured the formation of stoichiometric γ-Ta$_2$N, and indeed the position of the XRD peak shifted to its theoretical position, and the excess N was available to form the amorphous Si-N tissue between the crystalline phases. Finally, a further increase in N was responsible for the onset of formation of the cubic δ-TaN phase in TaSiN-13 and TaSiN-15. The incorporation of N into the cubic δ-TaN phase also meant that less N was available to react with Si; hence, the concentration of Si for TaSiN-13 and TaSiN-15 (10 at.% Si) was smaller than for TaSiN-7 and TaSiN-10 (containing 13 and 14 at.% Si respectively).

## 3.2. Room and high temperature nanoindentation

Room temperature hardness and elastic modulus of the TaSiN coatings as a function of nitrogen content are shown in Table II and in Fig. 3. Hardness and elastic modulus both increased with nitrogen content, with the hardness varying between 29 and 36 GPa, and the elastic modulus, between 210 and 255 GPa.

Table II. Room temperature mechanical properties of TaSiN coatings

| Sample | TaSiN-7 | TaSiN-10 | TaSiN-13 | TaSiN-15 |
|---|---|---|---|---|
| Hardness (GPa) | 29.3 ± 0.2 | 34.4 ± 0.3 | 35.5 ± 0.5 | 36.2 ± 0.4 |
| Reduced Modulus (GPa) | 210.0 ± 1.1 | 239.0 ± 1.6 | 260.0 ± 4.0 | 255.3 ± 1.0 |
| $K_{IC}$ (MPa×m$^{0.5}$) | 2.80 ± 0.11 | 3.76 ± 0.05 | 3.80 ± 0.04 | 3.95 ± 0.03 |

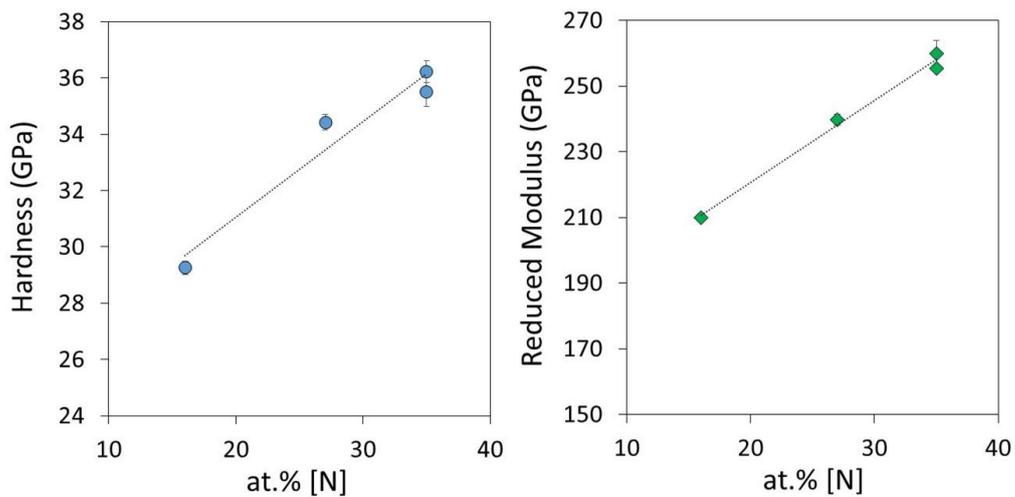

Fig. 3. Hardness and reduced modulus of deposited TaSiN coatings as a function of N content.

High temperature nanoindentation measurements were performed to investigate the effect of temperature on the coatings hardness and modulus. Figs. 4a and 4b show the hardness and reduced elastic modulus as a function of testing temperature, respectively. The effect of the testing temperature in the elastic modulus was not significant. However, the hardness reduction with temperature was similar for all coatings, with a drop of only 5% at 350°C and 15% at 500 °C. After the 500 °C tests ("after cooling values" in Fig. 4), the hardness values decreased between ~1.4 and 4.0 GPa, and the reduced moludi slightly increased, depending on the elemental composition of the coatings. The highest drop in hardness (~14%) and increase in modulus (~10%) was observed for sample TaSiN-7 and the lowest for TaSiN-15 (~4 % drop in hardness and ~4 % increase in modulus). This seems to indicate some microstructural change in the coatings, as will be discussed in more detail in section 3.4.

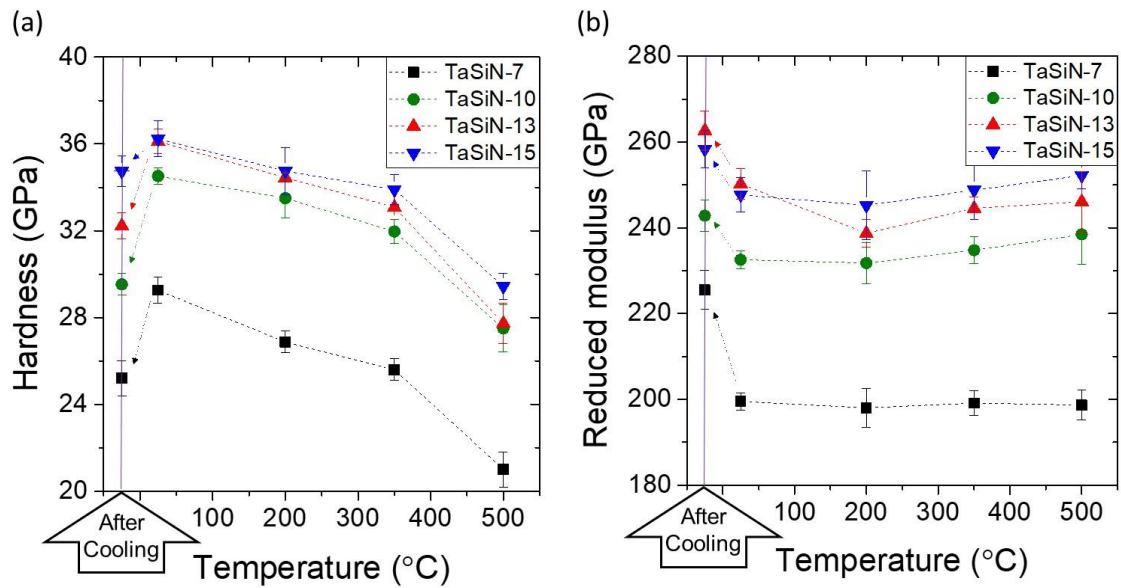

Fig. 4. (a) Hardness and (b) Reduced elastic modulus for the studied TaSiN coatings as a function of temperature. The hardness and modulus "after cooling" are also shown, being indicated by the large arrows in the "temperature" axis.

*3.3. Room and high temperature micropillar splitting results*

Hardness is not the only requirement for wear resistance; the coating toughness can be an equally important factor, particularly for abrasion, impact and erosive wear. In order to assess the coatings toughness, micropillar splitting tests were performed at the same temperature range as the nanoindentation tests, as described in section 2. A typical milled micropillar before fracture is shown in Fig. 5a, while post-fracture micropillar images are shown in Figs. 5b and 5c for TaSiN-

7 and TaSiN-15 coatings respectively. The load-displacement curves in Fig. 6a, and the post fracture micropillar images indicate that the deformation proceeded elasto-plastically in the initial stages until cracks were nucleated and propagated rapidly to the outer radius of the micropillar, triggering sudden splitting fracture (see Fig. 5c). The displacement-control tests were aborted manually after the critical load was detected by a sudden load drop and the feedback loop was unable to stop the tip crashing against the pillar base, right after the onset of the rapid cracking event. Sometimes, the fractured micropillar still stood after the test, typically for micropillars with a smaller aspect ratio that were loaded using a standard tip (rather than the special long shaft tip used for the high temperature tests), as observed in the inset of Fig. 5c. Despite the different post-fracture behaviour observed as a function of indenter tip, no variation in the calculated toughness values was observed for room temperature tests using either tip.

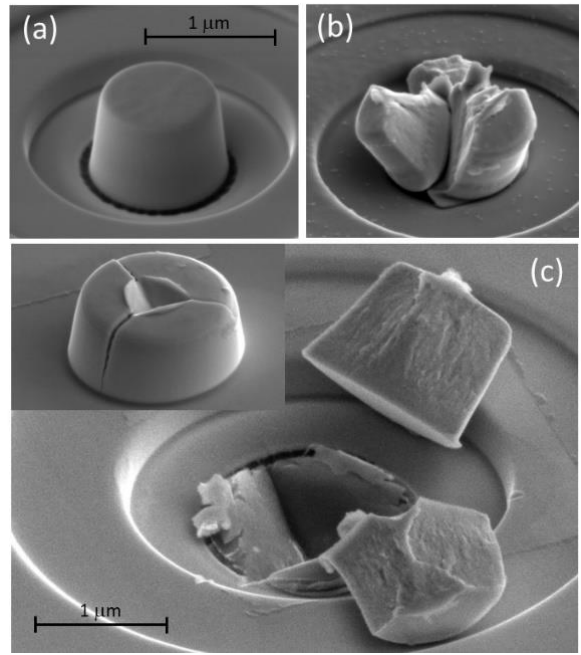

Fig. 5. SEM images of (a) micropillar before splitting test; (b) micropillar after splitting test at room temperature for TaSiN-7 and (c) micropillar after splitting test for TaSiN-15. The inset is a post-fracture micropillar tested with a standard tip. No variation in calculated toughness was observed for room temperature tests using either tip.

The fracture toughness was determined using equation (1), based on the critical load ($P_c$) at failure, the pillar radius $R$ and the dimensionless factor $\gamma$ determined from the E/H ratio [22]. The results are shown in table II and plotted in Fig. 7. Finally, representative load-displacement curves for micropillar splitting tests performed for all the coatings at 25, 350, 500 °C and after cooling are shown in Figs. 6a-d. The fracture toughness values estimated for all four coatings as a function of temperature are presented in Fig. 7. The toughness for all coatings increased with temperature, consistently with the hardness drop with temperature, which indicates enhanced plasticity with temperature. The increase in fracture toughness with temperature was particularly steep for TaSiN-7, to the point that no micropillar splitting took place for this coating at 500 °C. In this case, the tip penetrated continuously into the top of the micropillar while cracks extended from the corners of the indentation imprint, as shown in the inset of Fig. 6a, without inducing the splitting of the micropillar. Therefore, no fracture toughness could be determined for this coating at 500 °C. Finally, and contrary to the hardness changes observed, repeating the splitting tests after cooling down to room temperature (Figs. 6a-d) provided similar critical loads, and hence, similar toughness results to those obtained at room temperature before heating (Fig. 7).

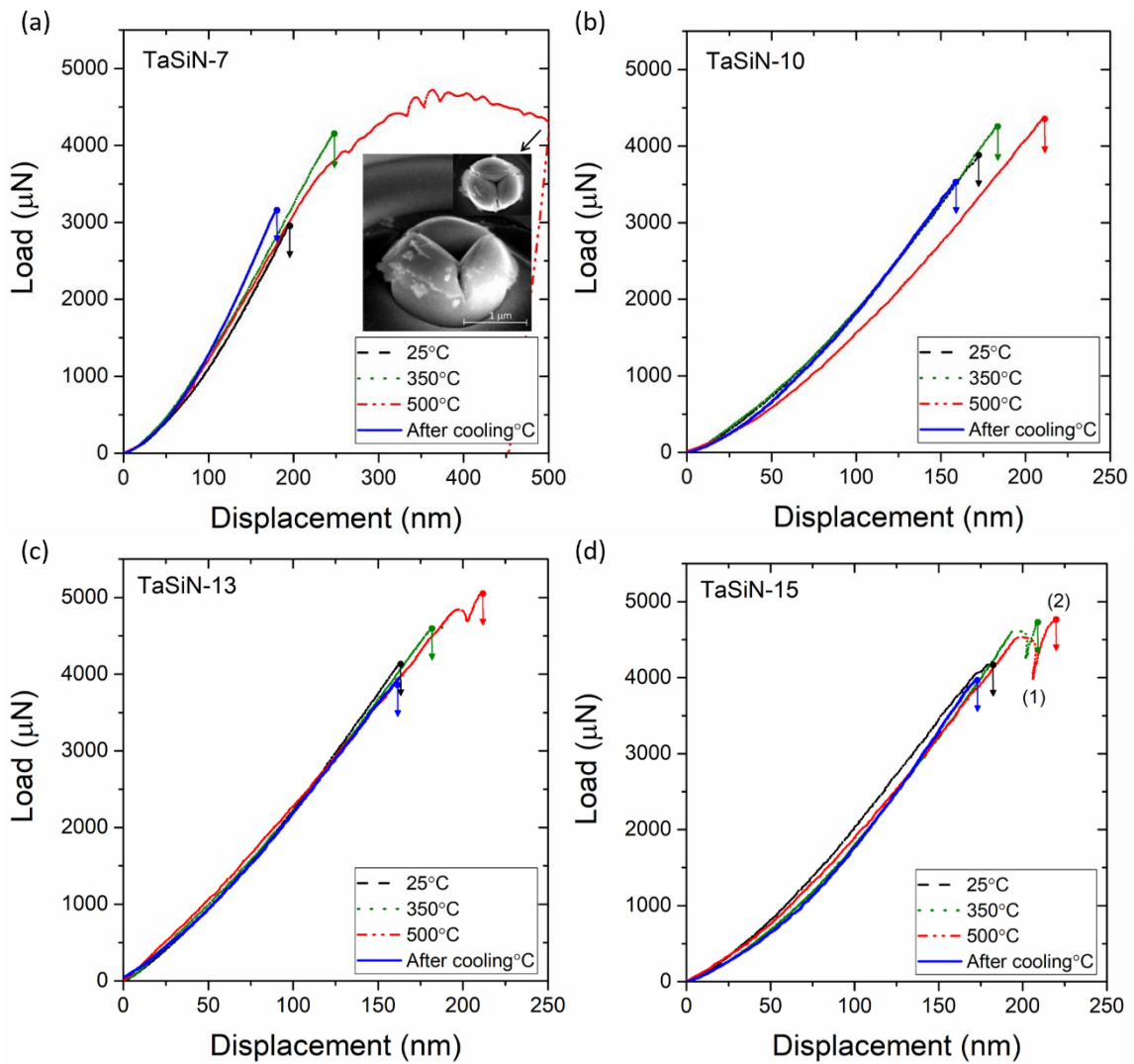

Fig. 6. Representative force-displacement curves showing critical loads for micropillar splitting tests performed at 25, 350, 500°C and after cooling for all four TaSiN coatings: (a) TaSiN-7 (the inset shows an SEM micrograph for TaSiN-7 after testing where no catastrophic fracture occurred), (b) TaSiN-10; (c) TaSiN-13; (d) TaSiN-15.

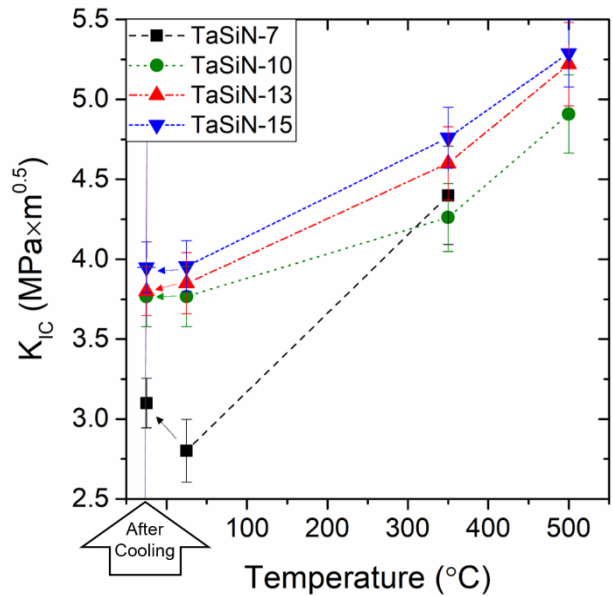

Fig. 7. Fracture toughness as a function of temperature for all TaSiN coatings.

Some of the TaSiN-13 and TaSiN-15 coatings tested at 350 and 500 °C exhibited a load drop (marked as (1) in the load-displacement curves of Fig. 6d) prior to the critical load. At this point, one of the tests was stopped for TaSiN-15 and the micropillar imaged by SEM, as shown in Fig. 8b. A crack can be observed at the top of the micropillar (inset of Fig. 8b) extending from one of the corners of the residual imprint to the outer radius of the micropillar. Figs 8a and 8c show post-fracture micropillars for TaSiN-15 tested at 350 and 500 °C respectively, which show how the micropillars adopted a mushroom shape in order to accommodate the pile-up formed around the indentation imprint, indicating a large increase in plasticity with temperature.

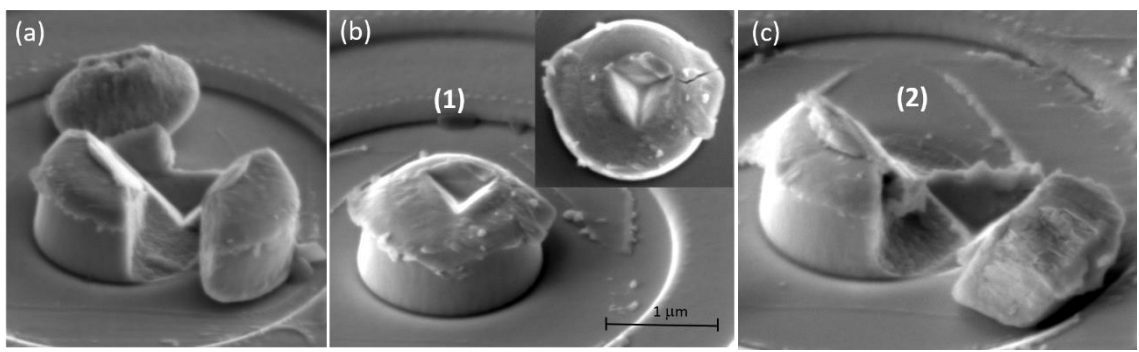

Fig. 8. SEM images of tested micropillars for the TaSiN-15 coating: (a) after splitting test performed at 350°C; (b) after splitting test at 500°C up to point (1) in Fig. 6(c) (the inset is a top

view of the same micropillar); and (c) after splitting test performed at 500°C - point (2) in Fig. 6(d), after total fracture.

*3.4. Microstructural changes during high temperature testing*

It is important to elucidate the potential microstructural changes that might occur during the high temperature nanoindentation and micropillar splitting tests. According to the XRD spectra of Fig. 9, the phase composition remained essentially unchanged for all samples after the indentation tests at 500 °C, so the coatings microstructure seemed to be stable up to this temperature.

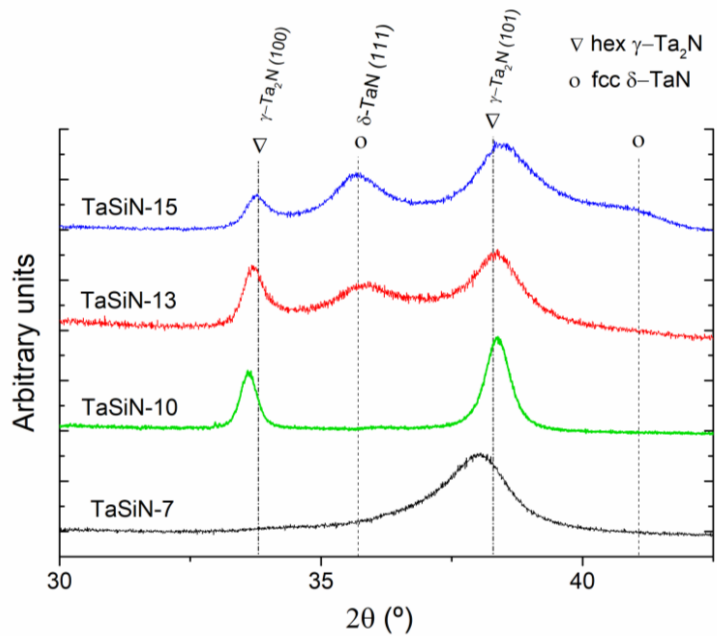

Fig. 9. XRD spectra of the TaSiN coatings after nanoindentation at 500 ºC.

Fig. 10 shows a cross-sectional TEM image of the TaSiN-7 coating after nanoindentation testing at 500 °C, where a distinct oxide surface layer with a thickness of ≈ 42 nm can be observed. This is not surprising since residual oxygen is always present (and subsequently uptaken during high temperature testing), even when the coatings were tested in a closed environment under a low flow of Ar. The oxide layer presented a multi-layered morphology. A compositional map for N and O across the top surface layer is shown in Fig. 11a. The oxide-rich surface layer was partially crystalline, as can be observed in the high-resolution TEM image of Fig. 10b, where the multilayer structure alternated amorphous and crystalline regions. The crystalline domains exhibited a lattice spacing of 0.235 nm, correlating to the period of hexagonal $Ta_2N$ (101) planes, which occupied ≈ 50% of the area, and were embedded in irregular Si-oxide dominated amorphous layers (see Fig. 10b). Apart from the surface oxide, no microstructural changes were observed for the rest of the coating thickness, which exhibited the same dense as-deposited nanocrystalline columnar microstructure, with lateral sizes of the order of 5-10 nm (see Fig. 11c). The selected-area diffraction (SAD) pattern for TaSiN-7 (inset of Fig. 10a) shows a discontinuous ring pattern, indicating that the coating had a noticeable texture, with the main reflection assigned to hexagonal γ-$Ta_2N$ (101), in agreement with XRD observations.

Similar observations were made with increasing the N content, with the only difference being the morphology and thickness of the surface oxide layer: the surface layer was more amorphous with a total thickness of only 15 nm for TaSiN-13 and TaSiN-15, as observed in Fig. 11b. Apart from the surface oxide layer, the microstructure of the coatings remained unchanged, showing columnar grains with an average grain size in the range 25-40 nm for TaSiN-15, composed of cubic δ-TaN and hexagonal γ-$Ta_2N$ phases, as revealed by the SAD patterns and in agreement with the XRD results, which were separated by amorphous regions, presumably Si-N rich (as pointed by the arrow in Fig. 11d).

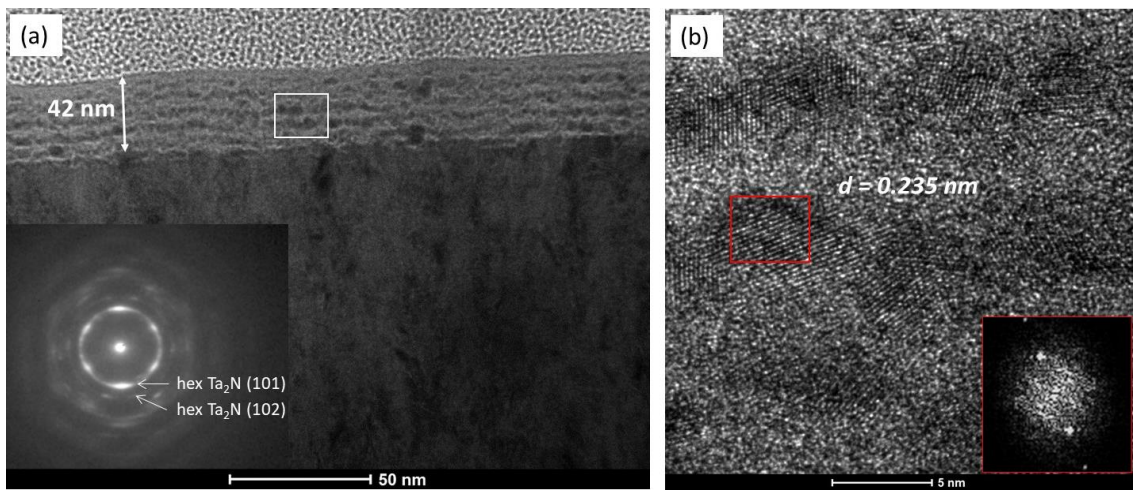

Fig. 10. (a) Cross-sectional bright-field (BF) TEM image of the TaSiN-7 coating. Inset is the selected-area diffraction (SAD) pattern of the area beneath the surface layer; (b) Cross-sectional high-resolution TEM image of the oxide-rich surface layer area enclosed by the white square in (a). The inset is a Fourier transform of the crystalline region delimited by the red square.

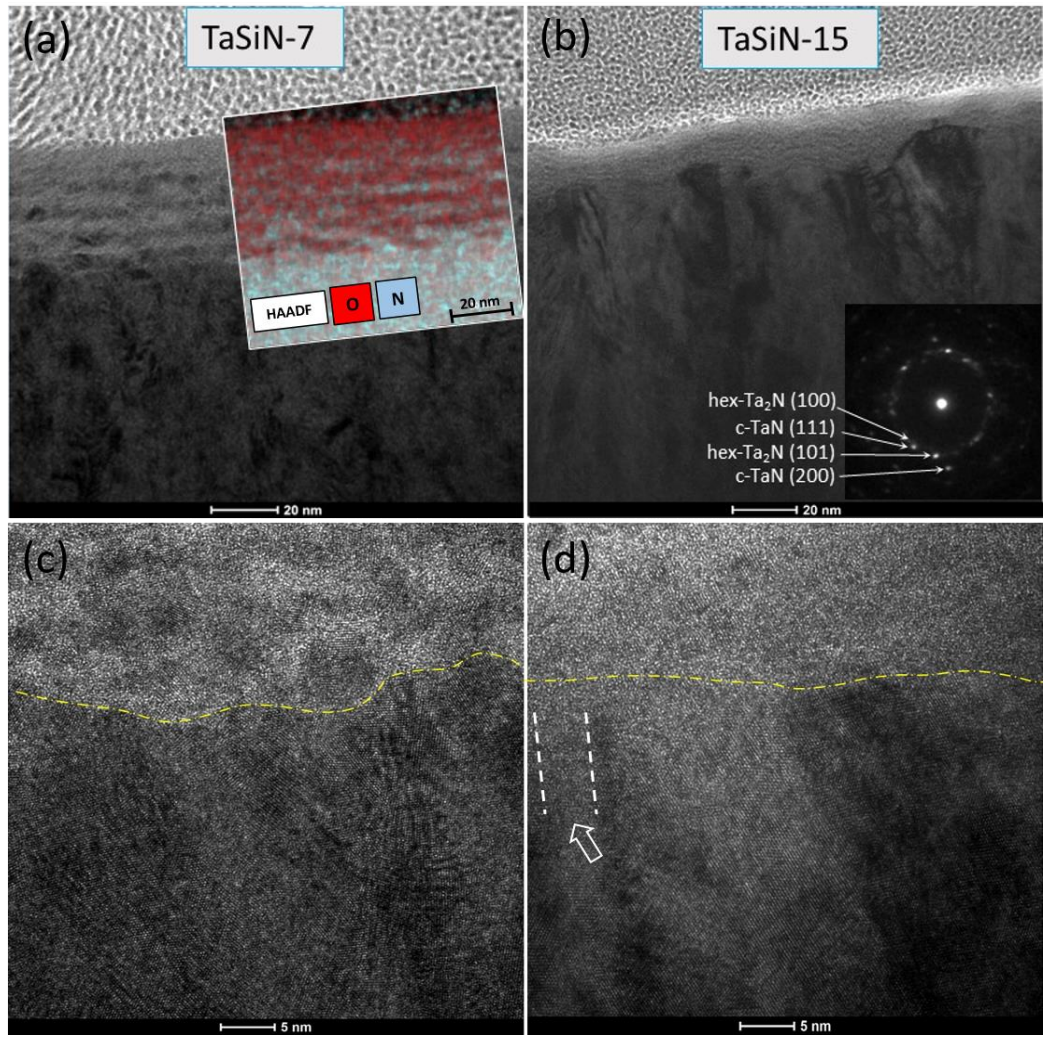

Fig. 11. (a) Cross-sectional BF TEM image of the TaSiN-7 coating. The inset is the EDX map of the near surface layer where red corresponds to O and blue to N; (b) Cross-sectional BF TEM image of the TaSiN-15. Inset is the SAD pattern; (c) and (d) are cross-sectional HR-TEM image of the TaSiN-7 and TaSiN-15 respectively. The dashed yellow line is the interface between the oxide-rich surface layer and the bulk of the coating; the white arrow in (d) points to an amorphous region.

4. Discussion

The hardness and elastic modulus of the TaSiN coatings were found to increase with nitrogen content. Hu *et al.* [29] showed that increasing the N content leads to a decrease in the fraction of Ta-Si bonds and to an increase in that of the Ta-N and Si-N bonds, which might be responsible for the increase in elastic modulus. Regarding hardness, our values were substantially higher than those reported by Nah *et al.* [12], Zeman *et al.* [8] and Chung *et al.* [30] for TaSiN films. The increase in hardness with N content was abrupt in the transition from TaSiN-7 to TaSiN-10, indicating that the formation of the amorphous Si-N tissue between columnar grains offers a strong obstacle to deformation [31]. The slight increase in hardness for higher N contents seems to correlate well with a higher degree of crystallization of the harder cubic δ-TaN hard phase. Another reason for the increase in hardness with N content could be related to the increase in residual compressive stresses found for similar TaSiN coatings [13], which was shown to increase from ~1 GPa to ~2 GPa, as $N_2$ partial pressure during deposition increased from 7 to 15% and attributed to the incorporation of N atoms at interstitial sites.

The room temperature fracture toughness of the coatings was higher than published results obtained from micropillar splitting tests on monolithic hard ceramic coatings, such as PVD-deposited TiN ($\approx$2.3 MPa×m$^{0.5}$) [22] and CrN ($\approx$3 MPa×m$^{0.5}$) [26], but comparable to nanolayered AlCrSiN coatings (between 3 and 4 MPa×m$^{0.5}$) [27]. For monolithic hard ceramic coatings, fracture toughness typically decreases with increasing hardness [28]. Interestingly, in our case, the room temperature fracture toughness increased with N content for the harder TaSiN coatings (TaSiN-10, TaSiN-13 and TaSiN-15). The optimum combination of hardness and fracture toughness was found for $Ta_{55}Si_{10}N_{35}$, with N and Si contents of 35 at.% and 10 at.%, respectively, with a hardness of 36 GPa and a fracture toughness of 3.95 MPa×m$^{0.5}$ at room temperature. This constitutes a further indication that the formation of the amorphous Si-N tissue, not only improves the hardness, and hence the resistance to plastic deformation, but also increases fracture toughness by delaying inter-columnar cracking leading to brittle fracture. This is consistent with results found in similar coatings, such as CrAlN with Si additions, where the formation of amorphous $Si_3N_4$ around the nitride crystallites was shown to delay inter-columnar cracking [34].

All coatings retained a good combination of hardness and fracture toughness at 500ºC, with the hardness decreasing by 15% and the fracture toughness increasing by 30%, due to enhanced plasticity of the coatings. No evidence of significant microstructural changes or grain size enlargement were found from the XRD and TEM studies after high temperature testing. Therefore, the coatings presented high temperature stability and the reported high temperature hardness and fracture toughness values were not a consequence of microstructural changes suffered during high temperature testing. On the contrary, the observed hardness trend as a function of temperature is attributed to thermally activated plastic deformation, as has been observed in other hard nitride coatings [19]. In our case, the hardness decreased by only $\approx$ 5% at 350ºC while a larger drop ($\approx$ 15%) was observed at 500 ºC. This increase in plasticity with temperature was even more evident on the micropillar splitting tests performed at 500 ºC, where the cube corner tip induced massive pile-ups, as observed in Fig. 8b.

Finally, the formation of a thin surface oxide layer was found in all the coatings after high temperature testing. N is replaced by O upon oxygen intake into the coating surface, probably forming $SiO_x$, since the Gibbs free energy of formation of $SiO_2$ is lower than that of $Ta_2O_5$

(−752.53 J/mol of $O_2$ for the former, against −663,57 J/mol of $O_2$ for the latter, at 700ºC) [32]. A possible diffusion path for oxygen might be inter-columnar boundaries. The formation of this $SiO_x$ surface layer is beneficial to passivate the TaSiN coatings and increases their oxidation resistance. However, the thickness and structure of the oxide layer changed with N content. For the lowest N content (TaSiN-7), the surface layer, with a thickness of 42 nm, was composed of amorphous $SiO_x$ rich layers intercalated with TaN-rich nanocrystalline domains. A similar high temperature oxidation process has been observed in other Si containing coatings, such as SiBCN [33], in which the preferential Si oxidation produced a surface layer composed of a $SiO_x$ amorphous matrix with a fine distribution of BN nanocrystals. Increasing the N content, produced a more compact and thinner $SiO_x$ rich layer, that was only ≈15 nm thick for TaSiN-15. Therefore, the oxidation resistance of the coatings increased with the N content. This might be because higher N contents in Si containing coatings encourages Si-N bonding that is more oxidation resistant than the Si-Si bonding, as has been proposed before [13].

Interestingly, the hardness after high temperature testing decreased slightly, between 15% for the lowest N content to 4% for the highest, while the elastic modulus increased approximately by the same amount. Considering the thermal stability of the coatings, this seems to be a consequence of the formation of the surface oxide layer. The latter explains the lowest hardness drop and modulus enhancement observed for the TaSiN-15 coating, for which the oxide-rich surface layer was the thinnest (≈15 nm), while the TaSiN-7, with the thickest oxide layer (≈40 nm), experienced the largest changes in hardness and modulus. On the contrary, the fracture toughness values after high temperature testing were similar to the as-deposited room temperature values, providing further evidence of the high temperature stability of the TaSiN coatings, and implying that the surface oxide layer does not influence the fracture toughness results.

5. **Conclusions**

In this work, the microstructure and the high temperature mechanical properties of magnetron sputtered TaSiN hard coatings have been evaluated. Coatings with the optimum composition were obtained by varying the nitrogen partial pressure during the sputtering process. XRD data showed that the TaSiN coatings exhibited a disordered nanocrystalline columnar microstructure. The nanocrystalline phases were mainly hexagonal γ-$Ta_2N$, with the presence of cubic δ-TaN for the coatings with the highest N content. Hardness, modulus and fracture toughness increased with N content, presumably due to the formation of an amorphous Si-N tissue at the grain boundaries. Higher N contents induced further hardening due to the increased presence of the hard cubic δ-TaN phase. The optimum N and Si contents were about 35 at.% and 10 at.%, respectively, for which the $Ta_{55}Si_{10}N_{35}$ coating had a hardness of 36 GPa and a fracture toughness of 3.95 MPa×$m^{0.5}$ at room temperature.

The high temperature mechanical properties of the TaSiN coatings have been measured for the first time. The coatings retained good mechanical properties at 500°C, with the $Ta_{55}Si_{10}N_{35}$ coating showing a hardness value of 30 GPa and a fracture toughness of 5.3 MPa×m$^{0.5}$ at 500°C. The room temperature hardness dropped by about 4% and the elastic modulus increased by the same amount for the $Ta_{55}Si_{10}N_{35}$ after testing at 500 °C, while the microstructure remained very stable. The reduced hardness and increased modulus after cooling was related to the effect of the surface oxide layer formed after the high temperature testing, whose thickness decreased with N content. In conclusion, TaSiN hard coatings represent a promising material for high-temperature applications in mechanical engineering.

**Acknowledgements**

RBS measurements were performed at the Centro Nacional de Aceleradores (CNA) of Seville. This investigation was supported by the European Research Council (ERC) under the European Union's Horizon 2020 research and innovation programme (Advanced Grant VIRMETAL, grant agreement No. 669141) and by the Regional Government of Madrid (IND2018/IND-9668). LW acknowledges support from project Nº FINE-KF-2019-02. A.Z. and E.I.M. acknowledge the support by the U.S. National Science Foundation under award NSF/CMMI DFREF-1335502.

**Data availability**

The raw/processed data required to reproduce these findings cannot be shared at this time due to technical or time limitations. It can be provided upon request.

**References**

[1]   L. Hultman, Thermal stability of nitride thin films, Vacuum. 57 (2000) 1–30. https://doi.org/10.1016/S0042-207X(00)00143-3.

[2]   A. Zaman, Microstructure and Mechanical Properties of TaN Thin Films Prepared by Reactive Magnetron Sputtering, (2017). https://doi.org/10.3390/coatings7120209.

[3]   R. Westergård, M. Bromark, M. Larsson, P. Hedenqvist, S. Hogmark, Mechanical and tribological characterization of DC magnetron sputtered tantalum nitride thin films, Surf. Coatings Technol. 97 (1997) 779–784. https://doi.org/https://doi.org/10.1016/S0257-8972(97)00338-1.

[4]   M. Alishahi, F. Mahboubi, S.M. Mousavi Khoie, M. Aparicio, E. Lopez-Elvira, J. Méndez, R. Gago, Structural properties and corrosion resistance of tantalum nitride coatings produced by reactive DC magnetron sputtering, RSC Adv. 6 (2016) 89061–89072. https://doi.org/10.1039/C6RA17869C.

[5]   T. Riekkinen, J. Molarius, T. Laurila, A. Nurmela, I. Suni, J.K. Kivilahti, Reactive sputter deposition and properties of TaxN thin films, Microelectron. Eng. 64 (2002) 289–297. https://doi.org/10.1016/S0167-9317(02)00801-8.

[6]   J.O. Olowolafe, I. Rau, K.M. Unruh, C.P. Swann, Z.S. Jawad, T. Alford, Effect of composition on thermal stability and electrical resistivity of Ta-Si-N films, Thin Solid Films. (2000). https://doi.org/10.1016/S0040-6090(99)01113-X.

[7]   Y. Chen, Y. Gao, L. Chang, Oxidation behavior of Ta e Si e N coatings, Surf. Coat. Technol. 332 (2017) 72–79. https://doi.org/10.1016/j.surfcoat.2017.09.087.

[8]   P.T. Zeman, J. Musil, R. Daniel, High-temperature oxidation resistance of Ta – Si – N films with a high Si content, 200 (2006) 4091–4096. https://doi.org/10.1016/j.surfcoat.2005.02.097.

[9]   Y. Chen, Y. Cheng, L. Chang, T. Lu, Chemical inertness of Ta – Si – N coatings in glass molding, Thin Solid Films. 584 (2015) 66–71. https://doi.org/10.1016/j.tsf.2014.12.040.

[10]  Y.-I. Chen, K.-Y. Lin, H.-H. Wang, Y.-R. Cheng, Characterization of Ta–Si–N coatings prepared using direct current magnetron co-sputtering, Appl. Surf. Sci. 305 (2014) 805–816. https://doi.org/https://doi.org/10.1016/j.apsusc.2014.04.011.

[11]  C. Shin, Y. Kim, D. Gall, J.E. Greene, I. Petrov, Phase composition and microstructure of polycrystalline and epitaxial TaN x layers grown on oxidized Si ( 001 ) and MgO ( 001 ) by reactive magnetron sputter deposition, 402 (2002) 172–182.

[12]  J.W. Nah, S.K. Hwang, C.M. Lee, Development of a complex heat resistant hard coating based on ( Ta , Si ) N by reactive sputtering, 62 (2000) 115–121.

[13]  A. Zaman, Y. Shen, E.I. Meletis, Microstructure and Mechanical Property Investigation of TaSiN Thin Films Deposited by Reactive Magnetron Sputtering, Coatings. 9 (2019) 338. https://doi.org/10.3390/coatings9050338.

[14]  K. Bobzin, High-performance coatings for cutting tools, CIRP J. Manuf. Sci. Technol. 18 (2017) 1–9. https://doi.org/10.1016/j.cirpj.2016.11.004.

[15]  A.A. Adjaottor, E. Ma, E.I. Meletis, On the mechanism of intensified plasma-assisted processing, Surf. Coatings Technol. 89 (1997) 197–203. https://doi.org/10.1016/S0257-

8972(96)02893-9.

[16] M.N. J.R. Tesmer, Handbook of Modern Ion Beam Materials Analysis, MRS, 1995.

[17] E. Andrzejewska, R. Gonzalez-Arrabal, D. Borsa, D.O. Boerma, Study of the phases of iron-nitride with a stoichiometry near to FeN, Nucl. INSTRUMENTS METHODS Phys. Res. Sect. B-BEAM Interact. WITH Mater. ATOMS. 249 (2006) 838–842. https://doi.org/10.1016/j.nimb.2006.03.150.

[18] M. Mayer, SIMNRA, v 6.06, Max Plank Institut fur Plasmaphysik, (n.d.).

[19] S.J. Bull, J. Chen, S.J. Bull, On the factors affecting the critical indenter penetration for measurement of coating hardness On the factors affecting the critical indenter penetration for measurement of coating hardness, Vaccum. 83 (2009) 911–920. https://doi.org/10.1016/j.vacuum.2008.11.007.

[20] W.C. Oliver, G.M. Pharr, An improved technique for determining hardness and elastic modulus using load and displacement sensing indentation experiments, J. Mater. Res. 7 (1992) 1564–1583. https://doi.org/DOI: 10.1557/JMR.1992.1564.

[21] M. Sebastiani, K.E. Johanns, E.G. Herbert, G.M. Pharr, Measurement of fracture toughness by nanoindentation methods : Recent advances and future challenges, 19 (2015) 324–333. https://doi.org/10.1016/j.cossms.2015.04.003.

[22] M. Ghidelli, M. Sebastiani, K.E. Johanns, G.M. Pharr, Effects of indenter angle on micro-scale fracture toughness measurement by pillar splitting, (2017) 1–8. https://doi.org/10.1111/jace.15093.

[23] A.M. Korsunsky, M. Sebastiani, E. Bemporad, Surface & Coatings Technology Residual stress evaluation at the micrometer scale : Analysis of thin coatings by FIB milling and digital image correlation, Surf. Coat. Technol. 205 (2010) 2393–2403. https://doi.org/10.1016/j.surfcoat.2010.09.033.

[24] N. Gordillo, R. Gonzalez-Arrabal, M.S. Martin-Gonzalez, J. Olivares, A. Rivera, F. Briones, F. Agullo-Lopez, D.O. Boerma, DC triode sputtering deposition and characterization of N-rich copper nitride thin films: Role of chemical composition, J. Cryst. Growth. 310 (2008) 4362–4367. https://doi.org/10.1016/j.jcrysgro.2008.07.051.

[25] C.. Ramberg, E. Blanquet, M. Pons, C. Bernard, R. Madar, Application of equilibrium thermodynamics to the development of diffusion barriers for copper metallization (invited), Microelectron. Eng. 50 (2000) 357–368. https://doi.org/10.1016/S0167-9317(99)00303-2.

[26] J.P. Best, J. Zechner, J.M. Wheeler, R. Schoeppner, M. Morstein, J. Michler, Small-scale fracture toughness of ceramic thin films: the effects of specimen geometry, ion beam notching and high temperature on chromium nitride toughness evaluation, Philos. Mag. 96 (2016) 3552–3569. https://doi.org/10.1080/14786435.2016.1223891.

[27] J.P. Best, J. Wehrs, M. Polyakov, M. Morstein, J. Michler, Scripta Materialia High temperature fracture toughness of ceramic coatings evaluated using micro-pillar splitting, Scr. Mater. 162 (2019) 190–194. https://doi.org/10.1016/j.scriptamat.2018.11.013.

[28] G. Dehm, B.N. Jaya, R. Raghavan, C. Kirchlechner, Acta Materialia Overview on micro- and nanomechanical testing : New insights in interface plasticity and fracture at small

length scales, Acta Mater. 142 (2018) 248–282. https://doi.org/10.1016/j.actamat.2017.06.019.

[29] R. Hu, Influence of nitrogen content on the crystallization behavior of thin Ta – Si – N diffusion barriers, 468 (2004) 183–192. https://doi.org/10.1016/j.tsf.2004.04.026.

[30] C.K. Chung, P.J. Su, Material characterization and nanohardness measurement of nanostructured Ta – Si – N film, 189 (2004) 420–424. https://doi.org/10.1016/j.surfcoat.2004.08.043.

[31] J. Musil, Hard nanocomposite coatings: Thermal stability, oxidation resistance and toughness, Surf. Coatings Technol. 207 (2012) 50–65. https://doi.org/10.1016/J.SURFCOAT.2012.05.073.

[32] David R. Lide, CRC Handbook of Chemistry and Physics, 84th Edition, J. Am. Chem. Soc. 126 (2004) 1586. https://doi.org/10.1021/ja0336372.

[33] J. He, M. Zhang, J. Jiang, J. Vlček, P. Zeman, P. Steidl, E.I. Meletis, Microstructure characterization of high-temperature, oxidation-resistant Si-B-C-N films, Thin Solid Films. 542 (2013) 167–173. https://doi.org/10.1016/J.TSF.2013.07.013.

[34] S. Liu, J. Wheeler, C. Davis, W. (Bill) Clegg, X. Zeng, The effect of Si content on the fracture toughness of CrAlN/Si3N4 coatings, J. Appl. Phys. 119 (2016) 25305. https://doi.org/10.1063/1.4939758.